# Forecasting and Analysis of Solar Energetic Particle Radiation Storms

Olga Malandraki[1], Michalis Karavolos[1], Arik Posner[2,3], Kostas Tziotziou[1], Henrik Droege[4], Bernd Heber[4] and Patrick Kuehl[4]

[1] National Observatory of Athens, IAASARS, Athens, Greece
[2] NASA Headquarters, USA
[3] NASA Johnson Space Center, SRAG, Houston, USA
[4] Christian-Albrechts-Universität zu Kiel, Germany
`omaland@noa.gr`

**Abstract.** Solar Energetic Particle (SEPs) with energies ranging from tens of keV to a few GeV, are a significant component in the description of the space environment. In this work, the HESPERIA REleASE product is emphasized, which, based on the Relativistic Electron Alert System for Exploration (REleASE) forecasting scheme, generates real-time predictions of the proton flux (30-50 MeV) at L1, making use of relativistic and near-relativistic electron measurements by the SOHO/EPHIN and ACE/EPAM experiments, respectively. The HESPERIA REleASE Alert is a notification system based on the forecasts produced by the HESPERIA REleASE product and informs about the expected radiation impact in real-time using an illustration and a distribution system for registered users. We also present and discuss the Advance Warning Times derived for the compiled list of major SEP events successfully forecasted over the last 2.5 years during solar cycle 25 by HESPERIA REleASE and recent developments.

## 1 Introduction

Solar Energetic Particles (SEPs) are transient injections into the heliosphere of protons, electrons and heavy ions, ranging in energy from tens of keV up to relativistic energies. As well as posing a threat to modern technology that heavily relies on spacecraft and posing a major radiation hazard to astronauts, they can also constitute a threat to avionics and commercial aircraft in extreme circumstances [1,2]. A warning system is required to predict SEP occurrence to mitigate radiation exposure [3]. The SEP Real-Time Forecasting HESPERIA products have been developed under the HESPERIA H2020 project and since 2015 provide significant results concerning the prediction of SEP events. The real-time and highly accurate forecasts as well as the timely performance offered by the HESPERIA products have attracted the attention of various space organizations (e.g. NASA/CCMC, SRAG) and also led to the selection and their integration into the ESA Space Weather Service Network (SWESNET) (https://swe.ssa.esa.int/noa-hesperia-federated). HESPERIA REleASE (see top part of Table 2 in [4]), based on the Relativistic Electron Alert System for Exploration (REleASE) forecasting scheme [5], generates real-time predictions of the proton flux (30-50 MeV) at L1, making use of relativistic and near-relativistic electron measurements by the SOHO/EPHIN and ACE/EPAM experiments, respectively [2]. In this work, we present and discuss the Advance Warning Times (AWTs) derived for the compiled list of major SEP events successfully forecasted over the last 2.5 years during solar cycle 25 by HESPERIA REleASE and recent developments.

## 2  HESPERIA REleASE forecasting system

The greatest assets of the HESPERIA REleASE forecasting system are that there is no need for any prior solar flare (soft X-ray observations) to issue forecasts and that forecasts are also provided in the case of backside flares, which most other schemes cannot handle. We note that 25% of the Solar Proton Events (SPEs) events observed at Earth's orbit are due to backside solar events [6].

To further explore the capabilities of HESPERIA REleASE we have looked at the temporal characteristics of 20 significant SPEs that occurred between September 2021 and December 2023, namely events that the proton flux exceeds the 0.1 cm$^{-2}$ s$^{-1}$ sr$^{-1}$ MeV$^{-1}$ threshold value in either the P3 (15.8-39.8 MeV) or and P4 (28.2-50.1 MeV) proton energy channel. The derived threshold crossing times as well as relevant proton alert times and onset times of the associated electron events are shown in Table 1.

| Date | Electron event Onset Time [in UT] | | Proton flux threshold crossing [in UT] | | Proton Alert Time [in UT] | | Advance Warning Time [in min] |
|---|---|---|---|---|---|---|---|
| | EPHIN | EPAM | P3 | P4 | EPHIN | EPAM | |
| 28/10/2021 | 15:54 | 16:24 | 18:08 | 17:55 | 16:08 | 17:20 | 107 |
| 20/01/2022 | 06:06 | - | 08:07 | 07:47 | 06:19 | | 88 |
| 28/03/2022 | 11:53 | 11:50 | 13:44 | 13:53 | 11:59 | 12:00 | 105 |
| 02/04/2022 | - | 13:39 | 14:38 | 14:39 | 14:03 | 13:39 | 35 |
| 09/07/2022 | 13:51 | 14:04 | 15:22 | 15:23 | 14:15 | - | 6 |
| 27/08/2022 | 03:20 | - | 10:32 | 13:52 | 10:30 | - | 2 |
| 13/01/2023 | 00:13 | - | 04:36 | - | - | - | - |
| 25/02/2023 | - | 19:50 | 21:26 | 21:29 | 19:58 | 20:00 | 88 |
| 13/03/2023 | 04:28 | - | 11:19 | 14:04 | - | - | - |
| 14/03/2023 | - | 08:30 | 10:44 | 15:13 | 09:48 | 10:24 | 56 |
| 23/04/2023 | 16:58 | 17:05 | 17:28 | 17:44 | 17:02 | - | 26 |
| 08/05/2023 | - | 09:05 | 13:12 | 07:23 | - | - | - |
| 09/05/2023 | 22:15 | 20:00 | 22:24 | 22:52 | 22:17 | 22:24 | 7 |
| 17/07/2023 | 18/07/2023 00:00 | 23:40 | 18/07/2023 01:06 | 18/07/2023 01:12 | 18/07/2023 00:06 | 23:54 | 60 |
| 28/07/2023 | 16:24 | 16:19 | 18:58 | 19:23 | 17:07 | 17:09 | 111 |
| 05/08/2023 | 07:42 | 07:40 | 10:03 | 10:14 | - | 09:59 | 4 |
| 05/08/2023 | 22:27 | 22:25 | 23:34 | 23:34 | 22:34 | - | 60 |
| 07/08/2023 | 21:32 | 21:00 | 22:09 | 22:19 | 21:33 | - | 36 |
| 01/09/2023 | 03:31 | 03:30 | 05:05 | 05:05 | 03:33 | 03:34 | 92 |
| 15/12/2023 | - | - | 15:09 | 15:09 | 11:07 | - | 242 |

**Table 1.** List of all 20 identified significant Solar proton Events (SPEs) during the period of September 2021 to December 2023, including relevant threshold crossing times, proton alert times and their associated electron event onset times. The last column shows the advance warning time. Adapted from [7].

Of particular interest for SEP forecasting is the estimation of the advance warning time, shown in the last column of Table 2, that denotes the advance warning time (AWT) in minutes from the first alert issue time (based either on ACE/EPAM or SOHO/EPHIN data) to the first onset of the significant SPE in either of the two energy channels. In Fig. 1 we show the distribution of AWT for the 20 considered SPEs. AWTs can go up to two hours with a mean value of ~70 min highlighting the unprecedented forecasting capabilities of HESPERIA REleASE.

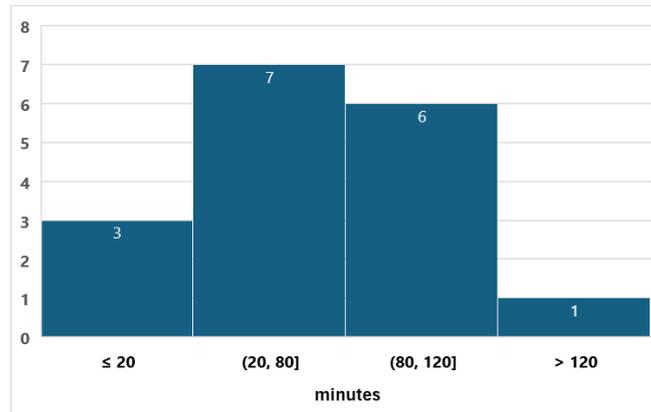

**Fig. 1.** Distribution of the Advance Warning Times (AWT), where applicable, for the SPEs presented in Table 1.

The efficacy of the HESPERIA REleASE forecasting tool in successfully predicting the SPE events is demonstrated in Figs. 2 and 3, that show, respectively, the events of October 28, 2021, and March 28, 2022. For the first event, an electron event is detected at 15:54 UT, triggering the HESPERIA REleASE tool to issue a proton alert at 16:08 UT with the SPE event observed to occur at 17:55 UT on the same day. For the second event the respective times are at 11:53 UT, 11:59 UT and 13:44 UT. Both events confirm the accuracy of the HESPERIA REleASE forecasting system, providing AWTs of 107 and 105 minutes, respectively. The clear temporal correlation between the electron event, the subsequent proton alert, and the SPE event highlights the system's capability to deliver timely and reliable predictions based on electrons as precursors.

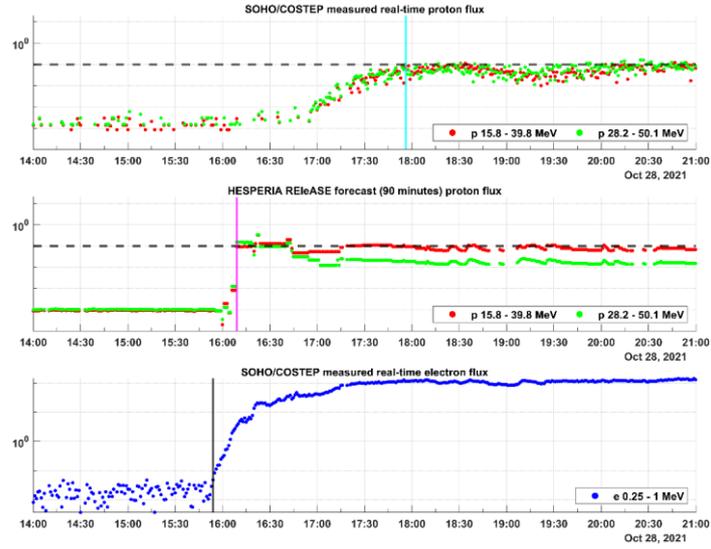

**Fig. 2.** The SPE event of October 28, 2021. The top panel presents the real-time proton fluxes also measured by SOHO/COSTEP, the mid panel the 60-min forecasted proton fluxes and bottom panel the real-time electron flux measured by SOHO/COSTEP. Black, magenta and cyan vertical lines indicate the onsets of the electron event, the proton alert issue time and the SPE occurrence time, respectively.

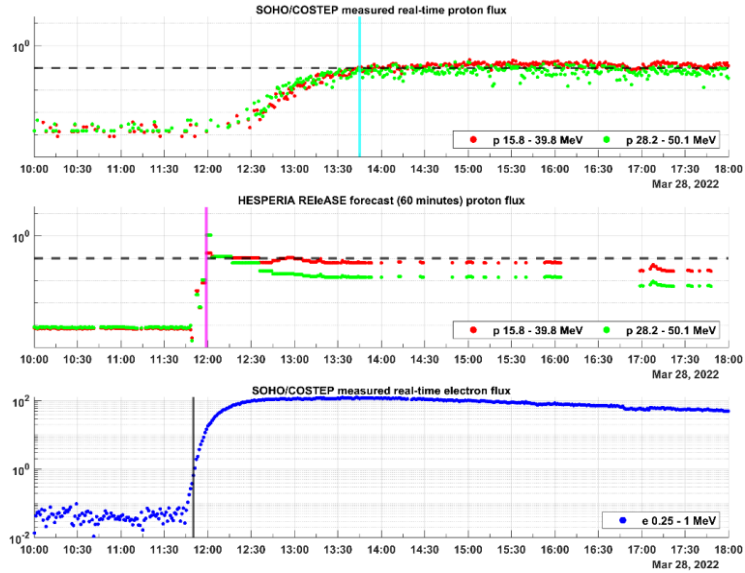

**Fig**. **3.** The SPE event of March 28, 2022. See caption of Fig. 2 for description of panels.

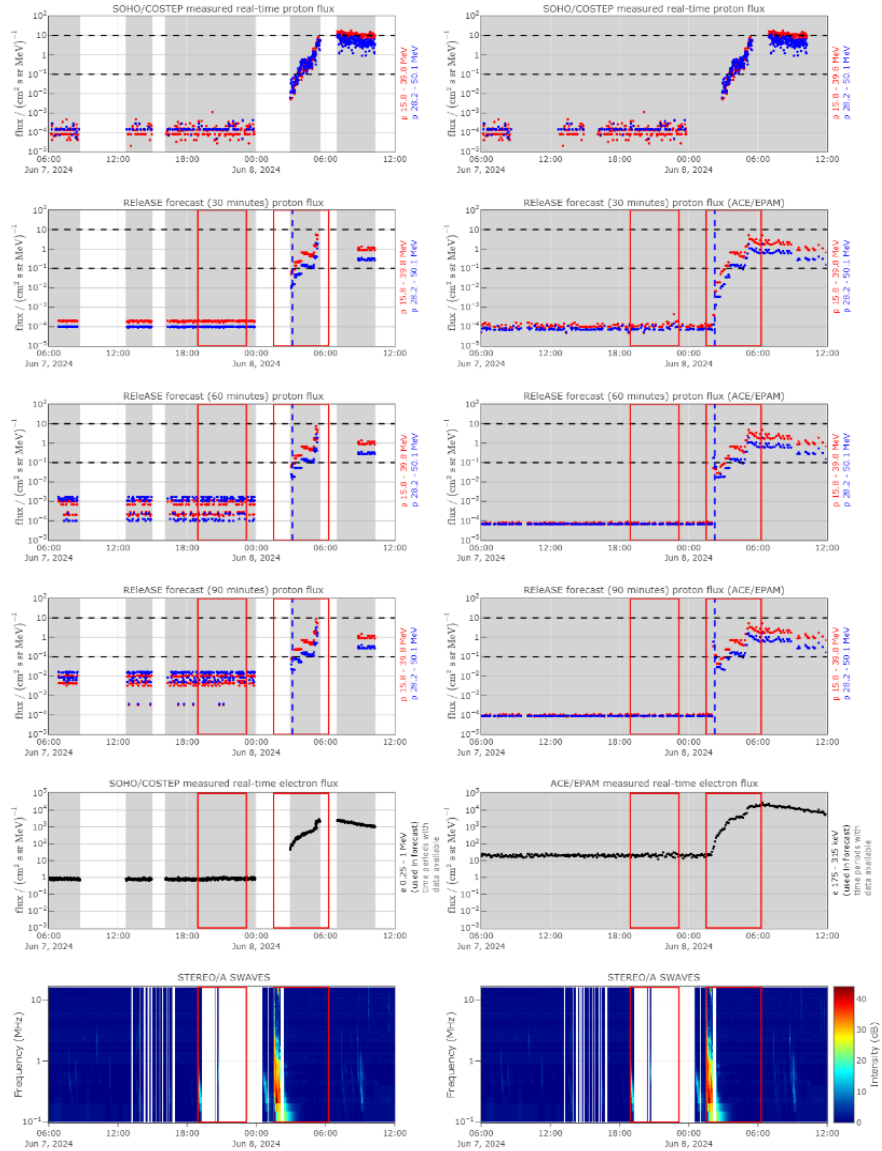

**Fig. 4.** The SPE event of June 8, 2024, as recorded by HESPERIA REleASE+. From top to bottom row panels show real-time proton fluxes, 30-min, 60-min and 90-min proton flux forecasts, real-time electron fluxes and real time STEREO/A SWAVES beacon radio fluxes. Vertical blue dashed lines indicate the time of the proton alert issue and red rectangles the forecasting window FW set by the radio module (see text).

A considerable update of HESPERIA REleASE is now available, namely HESPERIA REleASE+ that improves forecasting with evidence of particle escape from the Sun, using and qualifying type III radio bursts, associated with electron beams accelerated in solar eruptive events and observed by STEREO A/SWAVES, as a timely and reliable proxy of particle escape from the Sun onto open field lines [7]. To this end, a radio module has been built that automatically identifies any present type III radio burst in real-time STEREO A/SWAVES beacon data. This pairing of HESPERIA REleASE with relevant radio information is expected to substantially eliminate false alarms.

Fig. 4 shows a snapshot of the June 8, 2024 event as captured by HESPERIA REleASE+. The proton alert issue time (vertical blue dashed line) falls well within the forecasting window (FW, red rectangle) that denotes the maximum time interval (set by the radio module) within which a significant SPE is expected to occur after the onset of the qualified associated strong Type-III radio burst.

## 3   Summary

We have further explored the capabilities of the HESPERIA REleASE forecasting system and carried out a study of 20 significant SPE events, successfully forecasted between September 2021 and December 2023. We have found that the AWT for the considered SPEs can go up to two hours with a mean value of ~70 min. Thus, HESPERIA REleASE significantly improves mitigation of adverse effects in space from a significant solar radiation storm, providing ample time of forewarning to space weather users, which is also valuable for astronaut protection in the context of human exploration.